\documentclass[referee]{raa}            

\usepackage{graphicx,times}             
\newcommand{\npa}{Nucl. Phys. A }
\newcommand{\scie}{Science}

\newcommand{\cpl}{Chin. Phys. Lett. }
\newcommand{\aass}{Acta Astronom. Sin. Suppl. }
\newcommand{\epja}{Eur. Phys. J. A}

\begin{document}
   \title{Note on fast spinning neutron stars with unified equations of states
}

   \volnopage{Vol.0 (20xx) No.0, 000--000}      
   \setcounter{page}{1}          

   \author{A. Li
      \inst{1}
   \and N. B. Zhang
      \inst{2}
    \and B. Qi
      \inst{2}
   \and G. F. Burgio
      \inst{3}
   }

   \institute{Department of Astronomy, Xiamen University, Xiamen, Fujian 361005, China; {\it liang@xmu.edu.cn}\\
        \and
             Institute of Space Sciences, Shandong University, Weihai 264209, China\\
        \and
             INFN, Sezione di Catania, Via Santa Sofia 64, I-95123 Catania, Italy\\
   }

   \date{Received~~2018 month day; accepted~~2018~~month day}

\abstract{For the propose of confronting updated pulsar observations with developed neutron star equation of states (EoSs), we employ four unified EoSs for both the core and the crust, namely BCPM, BSk20, BSk21, Shen-TM1, as well as two non-unified EoSs widely used in the literature, i.e. APR and GM1 EoS, which are commonly matched with the Negele-Vautherin and the Baym-Pethick-Sutherland crust EoS. All the core EoSs satisfy the recent observational constraints of the two massive pulsars whose masses are precisely measured. We show that the NS mass-equatorial radius relations are slightly affected by the smoothness at the core-crust matching interface. Moreover, the uncertainties in the crust EoS and the matching interface bring insignificant changes, even at maximally rotating (Keplerian) configurations. We also find that for all four unified EoSs, rotation can increase the star's gravitational mass up to $18\%-19\%$ and the equatorial radius by $29\%-36\%$, which are consistent with the previous calculations using non-unified EoSs. For stars as heavy as 1.4 M$_{\odot}$, the radius increase is more pronounced, reaching $41\%-43\%$, i.e. $5-6$ km. Moreover, by confronting the results using unified EoSs, which give the correct empirical values at saturation density, with two controversial determinations of the radius for the fast rotator 4U 1608-52, we address that a small radius may be better justified for this source.
}

   \authorrunning{A. Li et al.}            
   \titlerunning{Note on fast spinning NSs with unified EoSs}  

   \maketitle

%
%
\section{Introduction}           
\label{sect:intro}

Neutron stars (NSs) are by far one of the most interesting observational objects, since many mysteries remain on them due to their complexity. Multi-messenger observations with advanced telescopes such as Advanced LIGO and VIRGO~(\cite{ligo}) , FAST~(\cite{fast}), SKA~(\cite{ska}),  NICER~(\cite{nicer}), HXMT~(\cite{hxmt}), eXTP~(\cite{extp}), AXTAR~(\cite{axtar}), will hopefully provide precise measurements of their mass and/or radius, thus improving our current knowledge of such stellar objects and their equation of states (EoSs). Those observations would also serve as a valuable guidance for improving models of such matter. Theoretically, a wide range of matter density from $\sim 0.1$ g cm$^{-3}$ in the star atmosphere, to values larger than $\sim 10^{14}$ g cm$^{-3}$ in the star core, is encountered in those objects. At the moment various model calculations, based on different theoretical frameworks, give distinct results especially for  the high-density inner crust and core regions. Accumulating observational studies might suffer from atmosphere modelling, burst cooling modelling and systematic data errors (\cite{atm16,mr16}). Therefore, for the study of NSs, particular attention should be paid on combining realistic enough theoretical models with accumulated NS observations. The present work is along this line.

Recently the importance of a unified EoS for the wide range of baryonic densities in a NS has been recognized (\cite{cp16,li16,li16prd, njl16,bcpm,bsk}). It may be necessary to calculate all EoS segments (outer crust, inner crust, liquid core) using the same nuclear interaction, since matching problems in non-unified EoS could bring nontrivial conflicts on the predictions of the stars' radii (\cite{cp16,bb2014}). The corresponding effects may be more important for distorted fast-rotating stars than for static ones (\cite{lrr03,qi16}). We examine in detail such problems in the present work.

The observations of massive NSs~(\cite{2mass16,2mass13,2mass10}) have already ruled out soft EoS which cannot reach 2 M$_{\odot}$ ($\rm M_\odot = 2 \times 10^{33} g$) maximum mass. This serves here as a criterion for our selection of NS (core) EoS. The calculations are mainly done for four available unified EoS, namely BCPM~(\cite{bcpm}), BSk20 and BSk21~(\cite{bsk}), Shen-TM1~(\cite{shen}). For comparison, we also use the non-unified APR~(\cite{apr}) and GM1~(\cite{gm1}) EoS for the core, with the
Baym-Pethick-Sutherland (BPS) ~(\cite{bps}) and the Negele-Vautherin (NV) (\cite{nv})
EoS for the outer and the inner crust, respectively.

The BCPM EoS, named after Barcelona-Catania-Paris-Madrid energy density functional~(\cite{bcpm}), is based on the microscopic Brueckner-Hartree-Fock (BHF) theory~(\cite{book}). The BSk20 and BSk21 EoS belong to the family of Skyrme nuclear effective forces derived by the Brussels-Montreal group~(\cite{bsk}). The high-density part of the BSk20 EoS is adjusted to fit the result of the neutron matter APR EoS~(\cite{apr}), whereas the high-density part of the BSk21 EoS is adjusted to the result of the BHF calculations using Argonne v18 potential plus a microscopic nucleonic three-body force. The Shen-TM1 EoS~(\cite{shen}) is based on a phenomenological nuclear relativistic mean field (RMF) model with TM1 parameter set, as well as the GM1 EoS which uses a different parameter set~(\cite{gm1}). The BPS outer crust EoS~(\cite{bps}) is based on a semi-empirical mass formula for matter from $10^7$ g cm$^{-3}$ to 3.4 $\times 10^{11}$ g cm$^{-3}$, whereas the NV inner crust EoS~(\cite{nv}) is based on quantal Hartree-Fock calculations for spherical Wigner-Seitz cells. In a previous study~(\cite{cp16}), the authors concluded that the largest uncertainties arise when the density dependence of the symmetry energies of the crust and the core matching EoS are very different. As we will see, the results are also dependent on the smoothness at the matching interface.

In addition, we are also interested in how fast a NS can rotate, and its maximum frequency, since very useful informations can be drawn from the spin frequency observations of a star, such as its possible evolution process~(\cite{Haensel16}), inner composition (e.g.,~\cite{Bhattacharyya16,li16,Steiner15,Li15a,Urbanec13,Read09,Worley08,xp07,Lattimer05,Morrison04}), possible central engine for short Gamma Ray Burst (SGRBs) (\cite{li16prd}), etc. For this purpose, the maximally allowed frequency, namely the Keplerian frequency, is calculated for the unified EoSs, as well as the corresponding stars' configurations.
In particular, we connect our calculations to radius observations for the X-ray burster in 4U 1608-52~(\cite{ozel16,atm16,4U1608-52,ozel10}). This NS has the largest observed spin frequency for an X-ray burster, and previously it was suggested to have a large radius ($> 14$ km) due to the flattening induced by the  fast rotation~(\cite{atm16,4U1608-52}). A further analysis gives a considerably different result, i.e. an initially smaller value of $8.3-10.3$ km~(\cite{ozel10}) was later updated to $6.8-12.6$ km~(\cite{ozel16}). The present study shows accurate theoretical calculations of NS matter with unified EoSs, as a result of which these two controversial determinations may be better understood.

The paper is organized as follows. We provide a short overview of the theoretical frameworks and discussions of our results in Sect.~2, before drawing conclusions in Sect.~3.

\section{Theoretical framework }
\label{sect:Theor}

\begin{figure}[t!]
{\centering
{\includegraphics[scale=0.85]{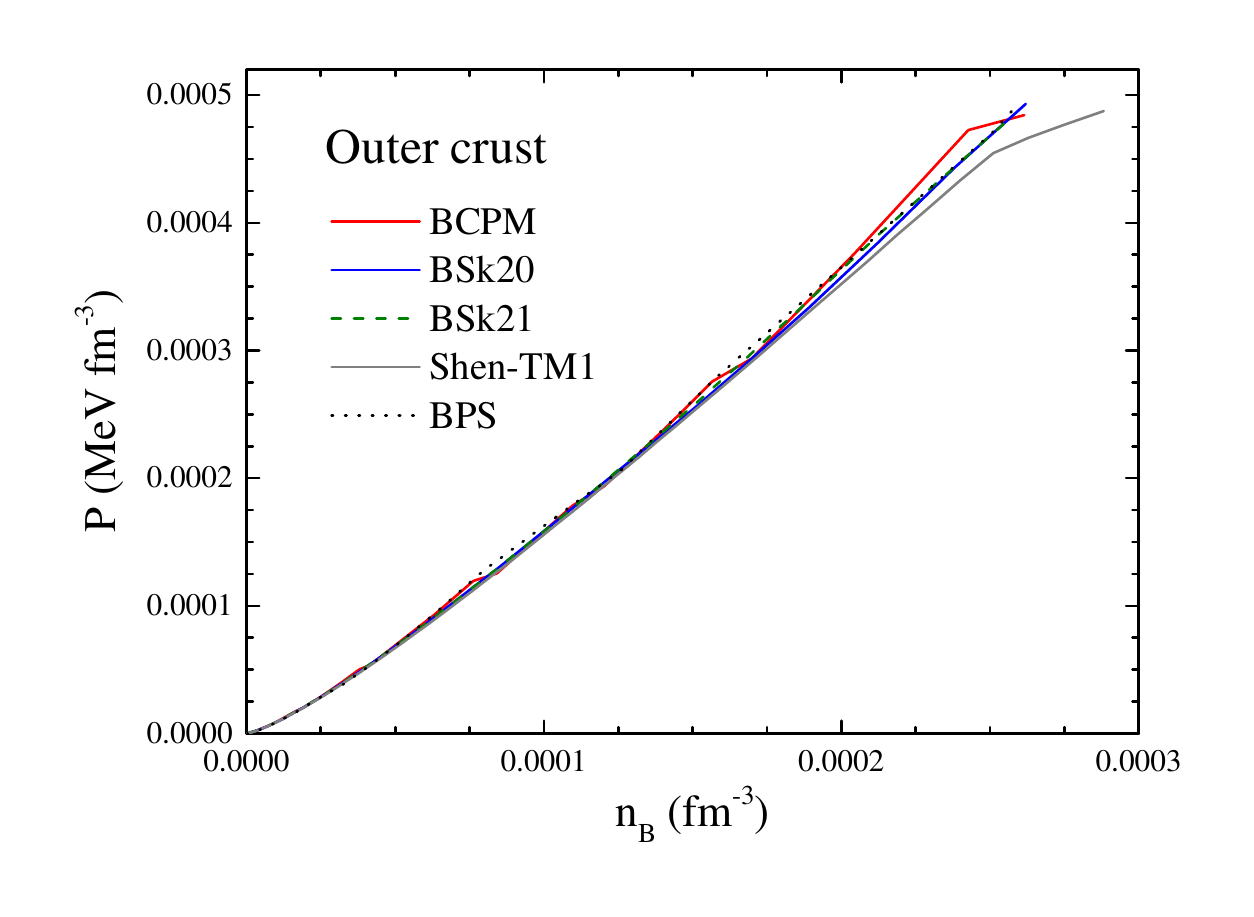}}
\par}
\caption{(Color online) Various EoSs for the outer crust employed in the present work. Among them, BCPM, Shen-TM1, BSk20, BSk21 are unified NS EoSs.
The BPS Eos is indicated by the black-dotted line.}
\label{fig1}
\end{figure}

\begin{figure}[t!]
{\centering
{\includegraphics[scale=0.85]{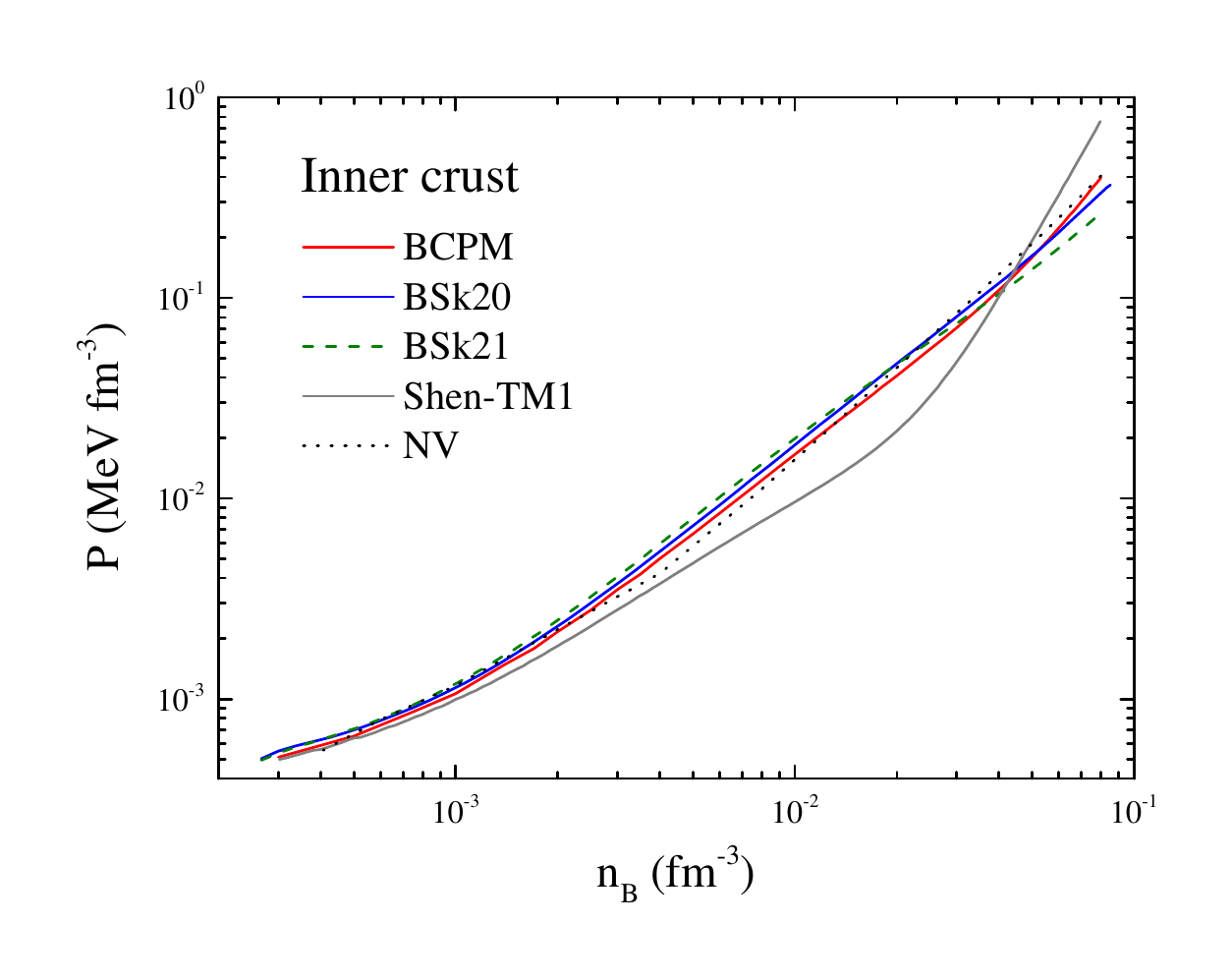}}
\par}
\caption{(Color online) Same as Fig.~1, but for the inner crust. The Negele-Vautherin (NV) EoS is indicated by a black dotted line. See text for details.}
\label{fig2}
\end{figure}

Theoretically, global properties like mass, radius, spin period of NSs are studied by using the overall EoSs as basic input, and ignoring their thin atmosphere ($\sim 0.1-10$) cm, where hot X-ray originates.

In the outer crust, at densities below $\sim 10^7$ g cm$^{-3}$, nuclei arrange themselves in a Coulomb lattice mainly populated by $^{56}$Fe nuclei. At higher densities ($10^7$ g cm$^{-3} - 4 \times 10^{11}$ g cm$^{-3}$) the nuclei are stabilized against $\beta$-decay by the filled Fermi sea of electrons, and become increasingly neutron-rich. The composition of the outer crust is mainly determined by the nuclear masses, which are experimentally measured close to stability, whereas the masses of very neutron-rich nuclei are not known, and they have to be calculated using nuclear models.

The inner crust is a non-uniform system of more exotic neutron-rich nuclei, degenerate electrons and superfluid neutrons. The density range extends from $\sim 4 \times 10^{11}$ g cm$^{-3}$ to nuclear saturation density $\rho_{\rm 0} = 2.8 \times 10^{14}$ g cm$^{-3}$, at which nuclei begin to dissolve and merge together. Non-spherical shapes of nuclear structures, generically known as nuclear ``pasta'', may appear at the bottom layers of the inner crust. Actually, one of NSs' irregular behaviours, the so-called $glitch$, is closely related to the inner crust EoS and the crust-core transition properties, see, e.g.,~\cite{li16,li16prd,li15,Piekarewicz14,crisis13,crisis12,Link99}.

\begin{table*}
\tabcolsep 1pt
\caption{EoS data at various crust-core matching interfaces.} \vspace*{-12pt}
\begin{center}
\def\temptablewidth{0.9999\textwidth}
{\rule{\temptablewidth}{0.5pt}}
\begin{tabular*}{\temptablewidth}{@{\extracolsep{\fill}}c  ccccc | cccccc}\hline
            {}
            &\multicolumn{5}{c}{inner crust}
            &\multicolumn{6}{c}{core} \\ \hline
            &{BCPM}&{Shen-TM1}&{BSk20}&{BSk21}&{NV}
            &{BCPM}&{Shen-TM1} &{BSk20}&{BSk21}&{APR} &{GM1} \\
\hline
             {$n$~(fm$^{-3})$}
            &{0.08}&{0.0797}&{0.0854} & 0.081 &{0.08}
            &{0.0825}&{0.08209} &{0.0864}& 0.0818 &{0.09} &{0.0918}   \\
             {$\rho$~(MeV fm$^{-3}$)  }
            &{75.93}&{75.43} &{81.06}& 76.75 & {75.95}
            &{78.30}&{77.76} &{82.08}& 77.7 &{85.36} &{82.74}   \\
             {$P$~(MeV fm$^{-3}$)}
            &{0.397}&{0.757} &{0.365} & 0.268 &{0.4058}
            &{0.4316}&{0.8293} &{0.3746}& 0.277 & {0.5793} &{0.8297}   \\\hline
\end{tabular*}\label{table2}
 {\rule{\temptablewidth}{0.5pt}}
\end{center}
\label{t:t1}
\end{table*}

In Fig.~\ref{fig1} we show the outer crust EoS for the different theoretical approaches discussed above. We observe that all outer crust EoSs display a similar pattern, with some differences around the densities where the composition changes from one nucleus to the next one. Only the Shen-TM1 EoS, based on a RMF model, shows a slightly different trend due to the semiclassical-type of masses calculations, in which $A$ and $Z$ vary in a continuous way, without jumps at the densities associated with a change of nucleus in the crust. On the other hand, the energy in the inner crust is largely determined by the properties of the neutron gas, hence the neutron matter EoS plays an important role. Moreover the treatment of complicated nuclear shapes, in a range of average baryon densities between the crust and the core, produces some uncertainties in the EoS of the inner crust, where some differences are visible. This is shown in Fig.~\ref{fig2}, where the well-known NV EoS is also displayed.

\begin{figure}[t!]
\vspace{0.3cm}
{\centering
{\includegraphics[scale=0.85]{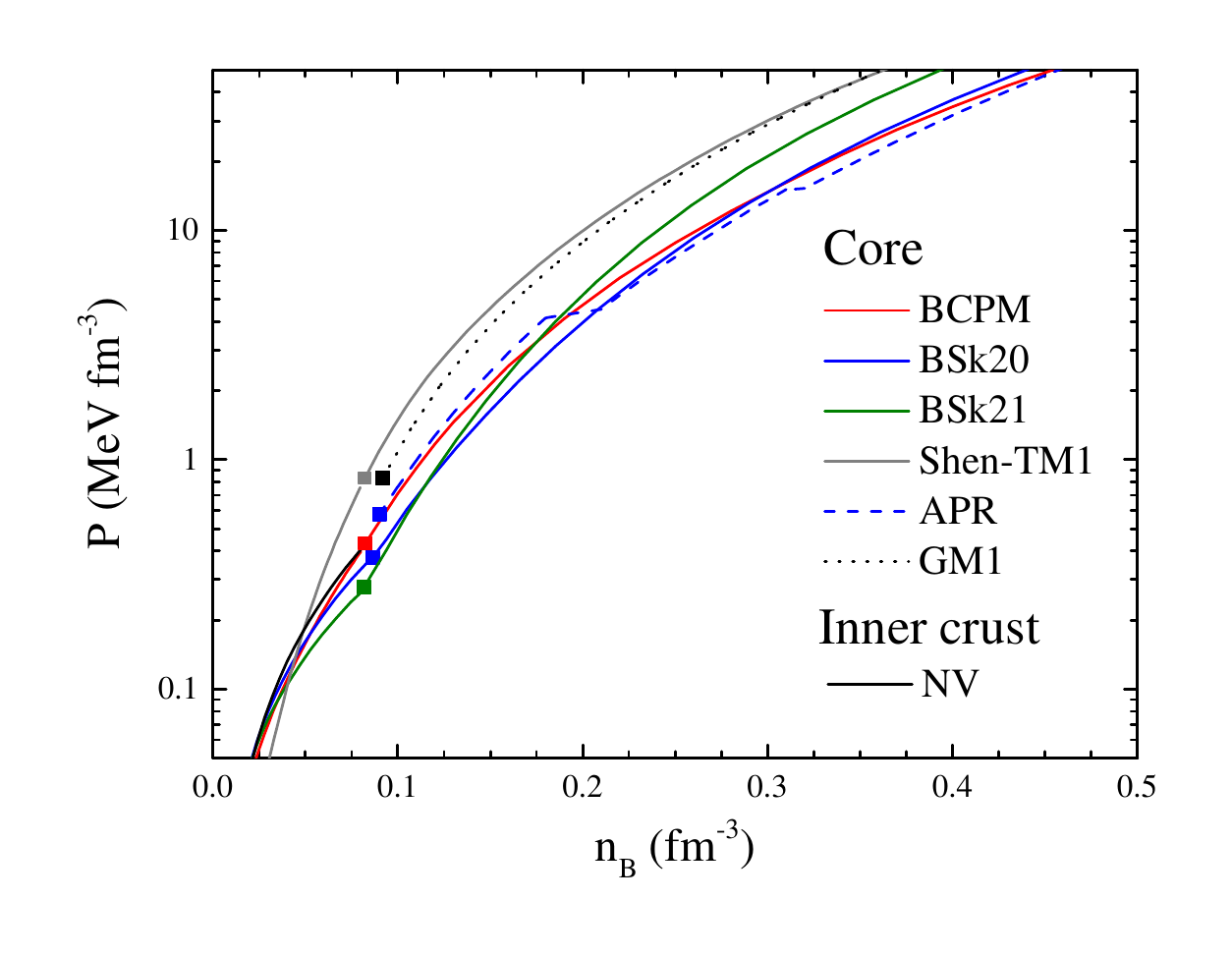}}
\par}
\caption{(Color online) The pressure is displayed vs. the density for various EoSs.}
\label{fig3}
\end{figure}

\begin{table*}
\tabcolsep 1pt
\caption{Properties of nuclear matter at saturation predicted by the EoSs employed in this study, in comparison with the empirical ranges. The number density $n_0$ is in fm$^{-3}$. The energy per baryon $E/A$ and the compressibility $K$ are in MeV, as well as the symmetry energy $E_{\rm sym}$ and its slope $L$ at saturation.} \vspace*{-12pt}
\begin{center}
\def\temptablewidth{0.9999\textwidth}
{\rule{\temptablewidth}{0.5pt}}
\begin{tabular*}{\temptablewidth}{@{\extracolsep{\fill}}cccccc}
\hline
~~&~$n_0$~&~$E/A$~&~$K$~&~$E_{\rm sym}$~&~$L$
\\
~EoS~&(fm$^{-3}$)&(MeV)&(MeV)&(MeV)&(MeV)
\\ \hline
 BCPM & 0.16 & -16.00 & 213.75 & 31.92 & 52.96
\\
Shen-TM1 & 0.145 & -16.26 & 281.14 & 36.89 & 110.79
\\
BSk20 & 0.159 & -16.08 & 241.4 & 30.0 & 37.4
\\
BSk21 & 0.158 & -16.05 & 245.8 & 30.0 & 46.6
\\
 APR  & 0.16 & -16.00 & 247.3  & 33.9 & 53.8
 \\
 GM1 & 0.153 & -16.32 & 299.2 & 32.4 & 93.9
  \\ \hline
 Empirical& $0.16\pm0.01$ & $-16.0\pm0.1$  & $220\pm30$ & $31\pm2$ & $\sim60\pm25$
\\ \hline
\end{tabular*}\label{table1}
 {\rule{\temptablewidth}{0.5pt}}
\end{center}
\label{tab2}
\end{table*}

\begin{figure}[t!]
\vspace{0.3cm}
{\centering
{\includegraphics[scale=0.85]{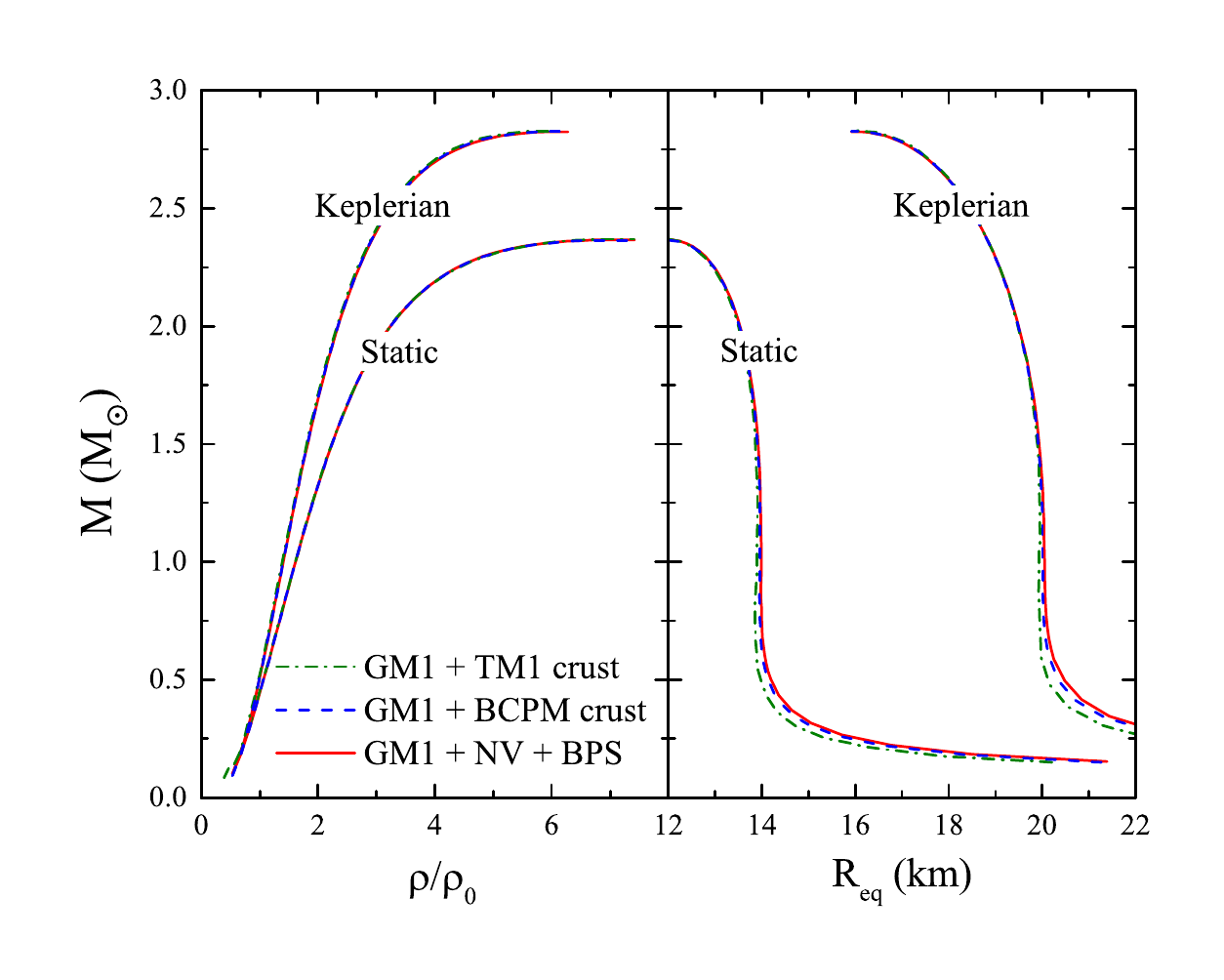}}
\par}
\caption{(Color online) NSs' masses as function of central energy density (left panel) and radius (right panel) for three cases of crust EoSs (Shen-TM1, BCPM, NV + BPS) matching with one GM1 core EoS, with the detailed EoS matching data shown in Table 1. The calculations are done for both static case and Keplerian rotating case. See text for details.}
\label{fig4}
\end{figure}

In Fig.~\ref{fig3} we show the above discussed EoSs, with the full symbols indicating the transition point from the inner crust to the core for each chosen EoS. Detailed EoS data at various crust-core matching interfaces are collected in Table \ref{t:t1}. We notice that APR and GM1 EoS have to be matched with an inner crust EoS,  at variance with BCPM, Shen-TM1, BSk20 and BSk21, and we actually did it by imposing that the pressure is an increasing function of the energy density. It is evident that the matching of GM1 core (dotted black) to Shen-TM1 crust (solid gray) shows a non-smooth behaviour for the $dP/d\rho$ (or $dP/dn$) slope, compared to the matching to BCPM and NV crust.  In the following we will study more in detail the corresponding effects.

Saturation properties of all core EoSs are presented in Table \ref{tab2}, with the empirical ranges listed in the last row. While ($n_0, E/A$) are very similar for all core EoS models, relatively large differences are present in the two RMF EoSs for ($K, E_{\rm sym}, L$). They are usually larger than the current empirical values, and this will give rise to larger star radii, as we will see later.

The strongest uncertainty in the EoS concerns the dense core part ($n_{\rm B}>(2-3)n_0$), which is mainly composed by uniform nuclear matter in  $\beta-$equilibrium with leptons. The determination of the EoS represents the main problem, also because first principle QCD calculations are difficult to perform in such a many-body system. Moreover, since in most of the model calculations available in the literature, a central density as high as $(7-10) n_0$  is found for the maximum mass, one or more types of strangeness phase transitions may take place in the NSs' innermost parts, for example, hyperons (\cite{Li14y,Li11y}),
kaon meson condensation (\cite{Li10k,Li06k,Li04k}), Delta excitation (\cite{Li16d}, quark deconfinement (\cite{Li15q,Li08q}). Due to our poor knowledge of the strange baryonic interaction and/or quark interaction, in this work we restrict ourselves to normal nuclear matter, leaving the effects from possible strangeness phase transitions to a future study.

The crust effects on the star's mass-radius relations in non-unified EoSs are shown in Fig.~\ref{fig4}, where three widely-used crust EoSs (Shen-TM1, BCPM, NV + BPS) are matched with one core EoS (GM1). Using tabulated EoS, we compute stationary and equilibrium sequences of rapidly rotating, relativistic stars in general relativity from the well-tested $rns$ code ($http://www.gravity.phys.uwm.edu/rns/$), assuming the matter comprising the star to be a perfect fluid. More details about the code can be found in~\cite{rns3,rns2,rns1}. It is clear that for both static case and Keplerian rotating case, the results hardly depend on how the inner crusts are described. This is true not only for the maximum mass and central densities, but also for the radii. For less massive stars, the crust-core matching has a slightly  larger effect on the radii, and the Shen-TM1 curve deviates a little from the other two, due to relatively larger difference in the crust-core interface for Shen-TM1 mentioned before. This deviation may be relevant only for  NSs' masses smaller than $1.0 \, M_{\odot}$ (\cite{bb2014}).

A complete set of results is shown in Table \ref{tab3} and Figs.~$5-6$. We notice that rotation increases both the gravitational mass and the radius, and at the same time it lowers the central density from $\sim 7 - 10 \rho_0$ to $\sim 6 - 9 \rho_0$.  In particular, rotation can increase the star's gravitational mass up to $\sim 18\% - 19\%$, and the star can be as massive as $\sim$ 2.61 M$_{\odot}$ in the  APR case. Also, the star becomes flattened and the corresponding circumferential radius is increased up to $\sim 3 - 4$ km, i.e. $\sim 29\% - 36\%$. For less heavy stars like 1.4 M$_{\odot}$, the radius increase is more pronounced, reaching $\sim 5 - 6$ km, i.e. $\sim 41\% - 43\%$. These results are consistent with the previous calculations using non-unified EoSs, for example as in \cite{rns3}. Possible highest spin frequencies $f_{\rm K}$ are all higher than 1000 Hz, while the current observed maximum is $f$ = 716 Hz~(\cite{716}) for PSR J1748-2446a in the globular cluster Terzan 5. A possible reason for this discrepancy is that the star fluid is suffering from $r$-mode instability. A simple estimation showed that this would lower the maximum frequency by $\sim 37\%$~(\cite{ZX}), which would satisfactorily explain the observations up to date.

\begin{table}
\centering
\caption{
Several properties of rotating NS for the selected EOSs:
Maximum gr
avitational mass,
corresponding central baryon density,
and Keplerian frequency.}
\setlength{\tabcolsep}{2pt}
\renewcommand{\arraystretch}{1.2}
\vspace{1mm}
\begin{tabular}{lcccccc}
\hline\hline
\multicolumn{2}{l}{EOS} & BCPM   & BSk20  & BSk21 & Shen-TM1 & APR*  \\
\hline
  & $M_{\rm max}/M_{\odot}$  & 1.98 & 2.17 & 2.28 & 2.18 & 2.20  \\
{Static}  & $R_{\rm eq}$(km) & 9.94 & 10.2 & 11.1  & 12.4 & 10.0 \\
  & $\rho_c/\rho_0$     & 10.9 & 10.0 & 8.35 & 7.08 & 10.1 \\
  \hline
  & $M_{\rm max}/M_{\odot}$  & 2.34 & 2.58 & 2.73 & 2.60 & 2.61  \\
{Keplerian}  & $R_{\rm eq}$(km) & 13.2 & 13.3 & 14.7 & 16.9 &13.0 \\
& $\rho_c/\rho_0$     & 9.44 & 8.80 & 7.34 & 5.94 &9.16  \\
  & $f_K ({\rm Hz})$   & 1791 & 1855 & 1661 & 1333 &1940  \\
\hline\hline
\end{tabular}
\label{tab3}
\end{table}

\begin{figure}[t!]
\vspace{0.3cm}
{\centering
{\includegraphics[scale=0.75]{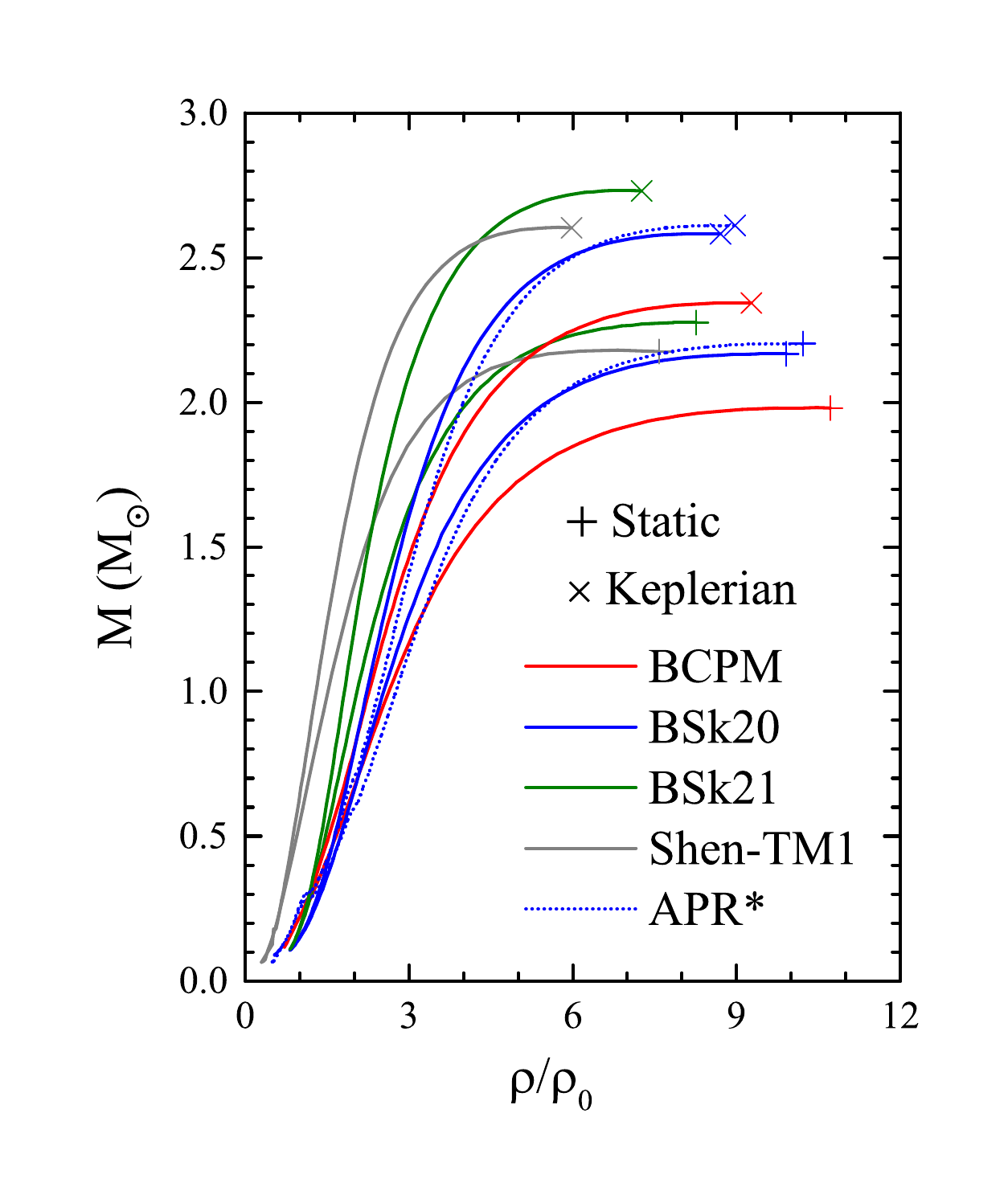}}
\par}
\caption{(Color online) NSs' masses as function of central density for four unified EoSs (BCPM, BSk20, BSk21, Shen-TM1) in solid lines. The results from one representative non-unified EoS APR*, namely ``APR + NV + BPS'', are also shown (dashed line). The calculations are done for both static case (lower curves) and Keplerian rotating case (upper curves) with plus (cross) symbols  labelling the corresponding maximum mass.}
\label{fig5}
\end{figure}

\begin{figure}[t!]
\vspace{0.3cm}
{\centering
{\includegraphics[scale=0.75]{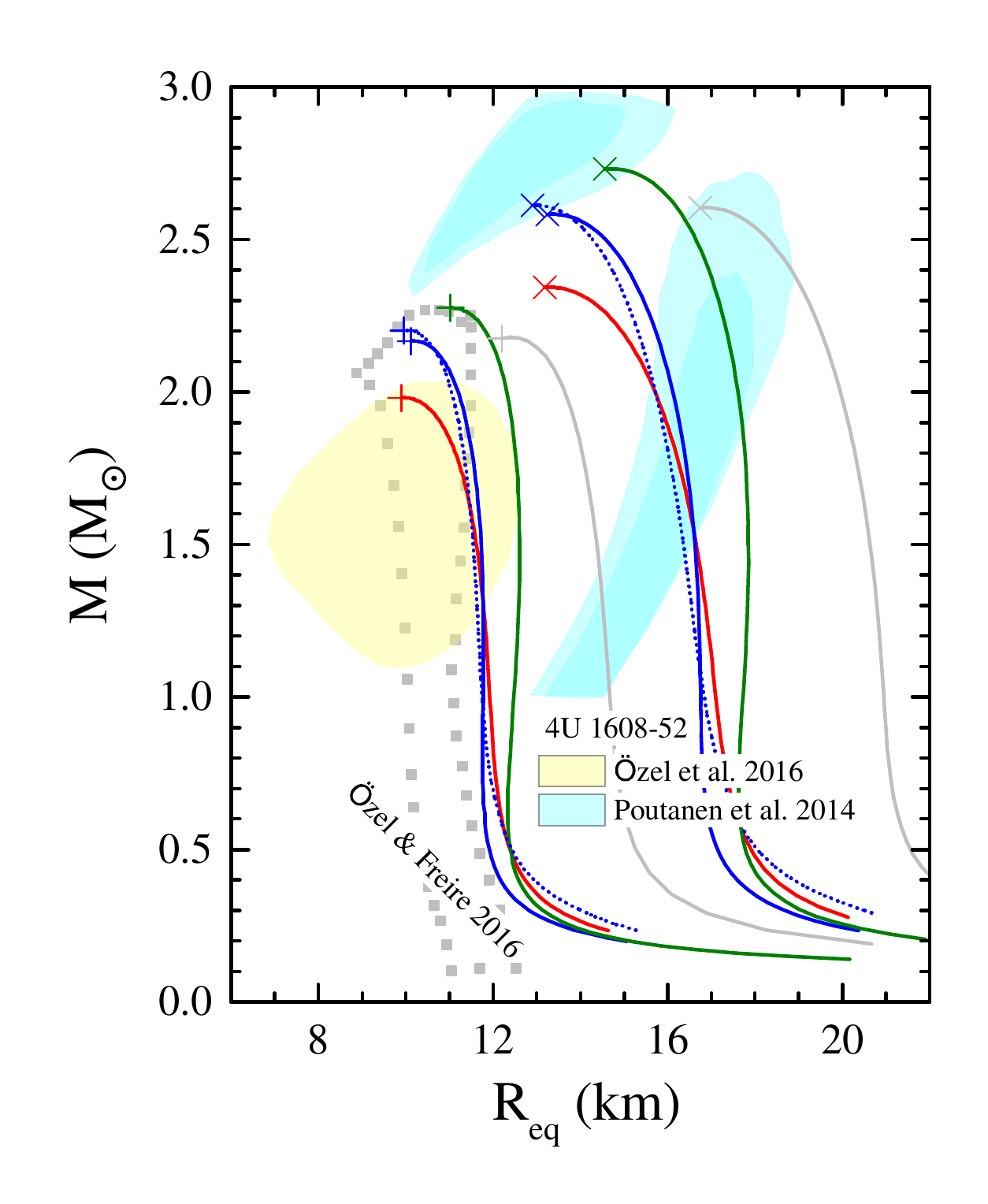}}
\par}
\caption{(Color online) Same as Fig.~\ref{fig5}, but NSs' masses are plotted as function of equatorial radius, together with observational determinations from a recent review~(\cite{ozel16rev}). Various results for the NS in the fastest-rotating X-ray burst (4U 1608-52) are also shown, in yellow shaded area~(\cite{ozel16,ozel10}) (at the $68\%$ confidence level) or blue shaded area~(\cite{4U1608-52}) (light/dark blue for the $68\%-90\%$ confidence level). See text for details.}
\label{fig6}
\end{figure}

In Fig.~\ref{fig6}, a recent review on radius determinations~(\cite{ozel16rev}) is also shown in the gray scattered areas, based on present combined constraints from the NSs in low-mass X-ray binaries (LMXBs) during quiescence and the NSs with thermonuclear bursts. We see that all EoS except the Shen-TM1 are in agreement with all current radius determinations. As discussed before, the reason for the much larger result for Shen-TM1 may be the very large values of  ($K, E_{\rm sym}, L$) compared to the other EoSs and to the empirical values. We then omit Shen-TM1 case in the following.

In addition, the constraints on the fastest-rotating X-ray burst 4U 1608-52, with a known spin at 620 Hz, are shown for comparison. They result from two analysis methods on different choices of photospheric radius expansion (PRE) bursts, shown in yellow shaded area~(\cite{ozel16,ozel10}) and blue shaded area~(\cite{4U1608-52}), respectively.

We have easily calculated that a spin frequency of 620 Hz can increase the radius only by less than $3\%$. For example in the BSk21 EoS, i.e. the stiffest among the unified EoS taken into account, the radius is lifted from 11.03 km to 11.29 km, corresponding to the maximum masses, respectively. Therefore present accurate calculations from various unified EoSs and the APR*  might reconcile better with \cite{ozel16,ozel10}, namely a smaller radius ($< 12.9$ km) is supported for the NS in 4U 1608-52 with the effects of the quadrupole moment and the ellipticity of the NS incorporated in the analysis~(\cite{ozel15}). In \cite{4U1608-52}, for typically assumed NS masses between 1.2 and 2.4 $\rm M_\odot$, the NS radius was strongly constrained to be above 13 km. The NS parameters were obtained assuming spherically-symmetric non-rotating NS, assuming an homogeneous distribution of the effective temperature, surface gravity and chemical composition in the atmosphere.  Rapid rotation can break the symmetry and change the NS shape and the observed flux. A preliminary study,
using a rotation-modified cooling tail method (\cite{juri16}),  indicates that for this source  the lower limit on the radius of the non-rotating NS may be reduced by as much as 10 per cent.

\section{Conclusions}

In the present work we aim to study how the (inner) crust EoS in non-unified EoSs affects the stars' rotating configurations as rapidly as the Keplerian limit, and also try to provide quantitative results for fast-rotating NSs.

For this purpose, four recently constructed unified EoSs, namely BCPM, BSk20, BSk21, and Shen-TM1, are employed to perform calculations of the fast rotating configurations based on the $rns$ code. The widely-used BPS outer crust EoS and the NV inner crust EoS are also used for comparison, as well as the APR and GM1 core EoS. All the core EoS chosen here satisfy the recent observational constraints of the two massive pulsars whose masses are precisely measured.

As far as the effects of the crust is concerning, we find that a non-smooth matching interface between core and inner crust produces a slight deviation in the mass-radius relation with respect to the unified description, even for the maximally-spinning (Keplerian) rotating configuration.
Those small changes are visible for star masses less than 1 $\rm M_\odot$, in agreement with previous findings (\cite{bb2014}).

For all the considered EoSs, rotational effects can increase the star's gravitational mass up to $\sim 19\%$, and  the corresponding circumferential radius up to $\sim 36\%$, depending on the core EoS. For stars as heavy as 1.4 M$_{\odot}$, the radius increase may reach up to $\rm \sim 5 - 6$ km. Moreover, by comparing the present calculations with recent simultaneous determinations of mass and radius for the fast rotator 4U 1608-52, we address that the NS is possibly rotating fast with a radius smaller than 13 km.

In terms of future methods, useful constraints could be possible by combining gravitational wave and electromagnetic observations from both coalescing NS-NS binaries and isolated NSs. New NS EoS consistent with both constrints from two-solar-mass and tidal deformability of GW170817 event~(\cite{ligo}) are in preparation (\cite{zhu18}). There has also been efforts to deduce information of interquark parameters from this event (\cite{zhou18}).

\begin{acknowledgements}
We would like to thank Z.-S. Li for valuable discussions. We appreciate H. Shen for providing us the Shen-TM1 EoS. We appreciate G.-Y. Shao for providing us the GM1 EoS. The work was supported by the National Natural Science Foundation of China (No. U1431107) and the Young Scholars Program of Shandong University, Weihai (Grant No. 2015WHWLJH01). Partial support comes from "NewCompStar", COST Action MP1304.
\end{acknowledgements}

\label{lastpage}

\end{document}